\documentclass{emulateapj}
\def \vt{\vartheta}
\def \farcs{\hbox{$.\!\!^{\prime\prime}$}}

\def\Cg{{\bf C}}
\def\d{{\rm d}}

\lefthead{Hoekstra et al.}
\righthead{First CFHTLS cosmic shear results}

\begin{document}

\title{First cosmic shear results from the Canada-France-Hawaii
Telescope Wide Synoptic Legacy Survey\footnotemark[$\dagger$]}

\footnotetext[$\dagger$]{Based on observations obtained with
MegaPrime equipped with MegaCam, a joint project of CFHT and
CEA/DAPNIA, at the Canada-France-Hawaii Telescope (CFHT) which is
operated by the National Research Council (NRC) of Canada, the
Institut National des Science de l'Univers of the Centre National
de la Recherche Scientifique (CNRS) of France, and the University
of Hawaii. This work is based in part on data products produced at
TERAPIX and the Canadian Astronomy Data Centre as part of the
Canada-France-Hawaii Telescope Legacy Survey, a collaborative
project of NRC and CNRS. }

\author{H.~Hoekstra\altaffilmark{1}, Y.~Mellier\altaffilmark{2,3},
L.~van Waerbeke\altaffilmark{4}, E.~Semboloni\altaffilmark{2}, 
L.~Fu\altaffilmark{2}, M.J.~Hudson\altaffilmark{5}, 
L.C.~Parker\altaffilmark{5}, I.~Tereno\altaffilmark{2,6}, 
\& K. Benabed\altaffilmark{2,3}}

\altaffiltext{1}{Department of Physics and Astronomy, University of Victoria,
Victoria, BC, V8P 5C2, Canada}

\altaffiltext{2}{Institut d'Astrophysique de Paris, 98bis Boulevard Arago,
75014 Paris, France}

\altaffiltext{3}{Observatoire de Paris, LERMA, 61 avenue de l'Observatoire,
75014 Paris, France}

\altaffiltext{4}{Department of Physics and Astronomy,
University of British Columbia, Vancouver, BC, V6T 1Z1, Canada}

\altaffiltext{5}{Department of Physics, University of Waterloo, Waterloo,
ON, N2L 3G1, Canada}

\altaffiltext{6}{Departamento de Fisica, Universidada de Lisboa, 1749-016
Lisboa, Portugal}

\begin{abstract}

We present the first measurements of the weak gravitational lensing
signal induced by the large scale mass distribution from data obtained
as part of the ongoing Canada-France-Hawaii Telescope Legacy Survey
(CFHTLS). The data used in this analysis are from the Wide Synoptic
Survey, which aims to image $\sim 170$ square degree in five
filters. We have analysed $\sim 22$ deg$^2$ (31 pointings) of $i'$
data spread over two of the three survey fields. These data are of
excellent quality and the results bode well for the remainder of the
survey: we do not detect a significant `B'-mode, suggesting that
residual systematics are negligible at the current level of
accuracy. Assuming a Cold Dark Matter model and marginalising over the
Hubble parameter $h\in[0.6,0.8]$, the source redshift distribution and
systematics, we constrain $\sigma_8$, the amplitude of the matter
power spectrum. At a fiducial matter density $\Omega_m=0.3$ we find
$\sigma_8=0.85\pm0.06$. This estimate is in excellent agreement with
previous studies. Combination of our results with those from the Deep
component of the CFHTLS enables us to place a constraint on a constant
equation of state for the dark energy, based on cosmic shear data
alone. We find that $w_0<-0.8$ at 68\% confidence.

\end{abstract}


\section{Introduction}

Weak gravitational lensing of distant galaxies by intervening massive
structures in the universe provides us with a unique, and unbiased way
to study the distribution of matter in the universe. Although weak
lensing has many applications in astronomy, from the study of galaxy
dark matter halos (e.g., Brainerd et al. 1996; Hudson et al. 1998;
McKay et al. 2001; Hoekstra et al. 2004) to galaxy clusters (e.g.,
Clowe et al. 1998; Dahle et al. 2002; Hoekstra et al. 2002a; Cypriano
et al. 2004), much recent work has been devoted to the measurement of
the statistical signal induced by large scale structure (a.k.a. cosmic
shear).

\begin{figure*}
\begin{center}
\leavevmode
\hbox{%
\epsfxsize=8.5cm
\epsffile{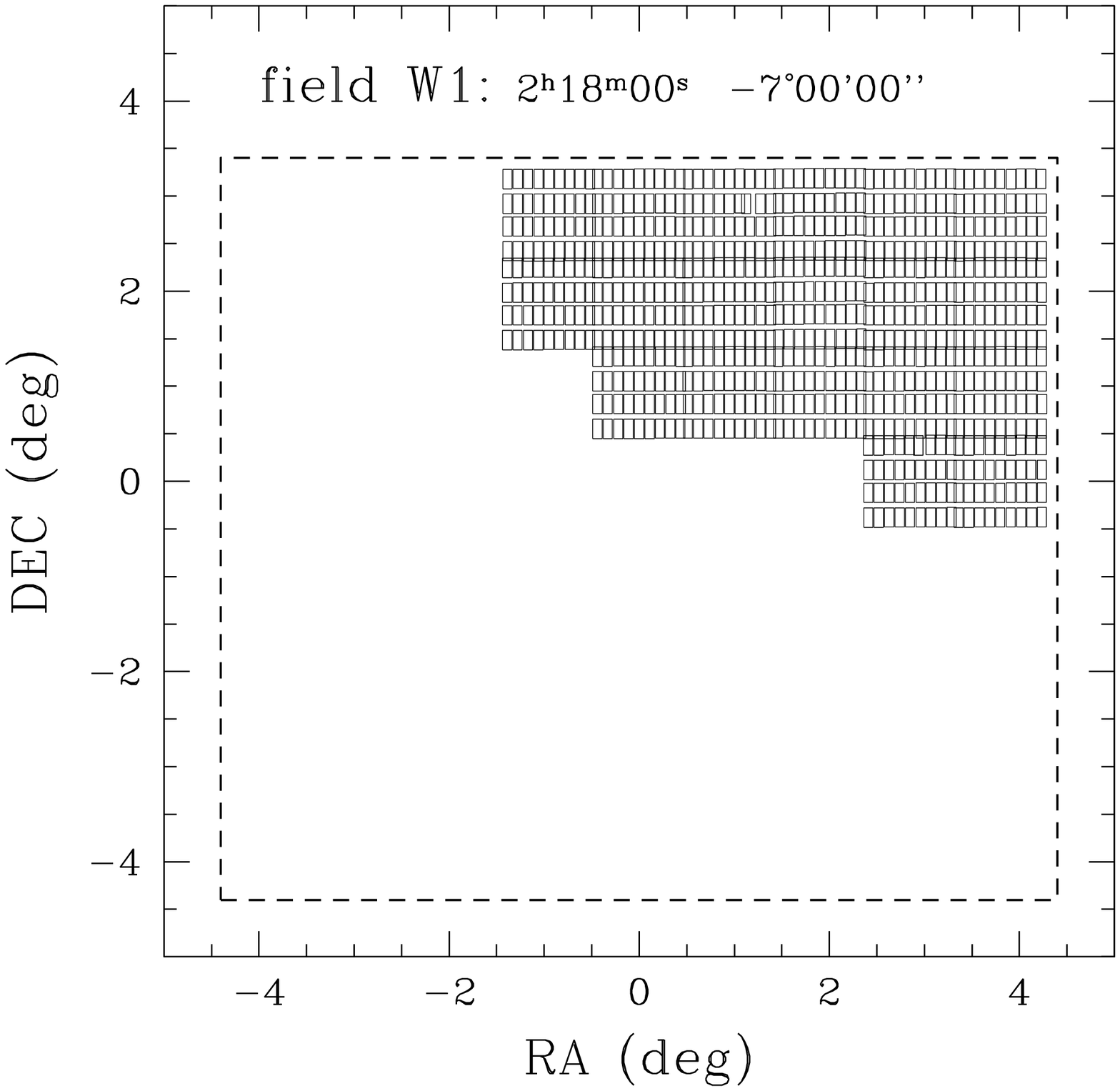}
\epsfxsize=8.5cm
\epsffile{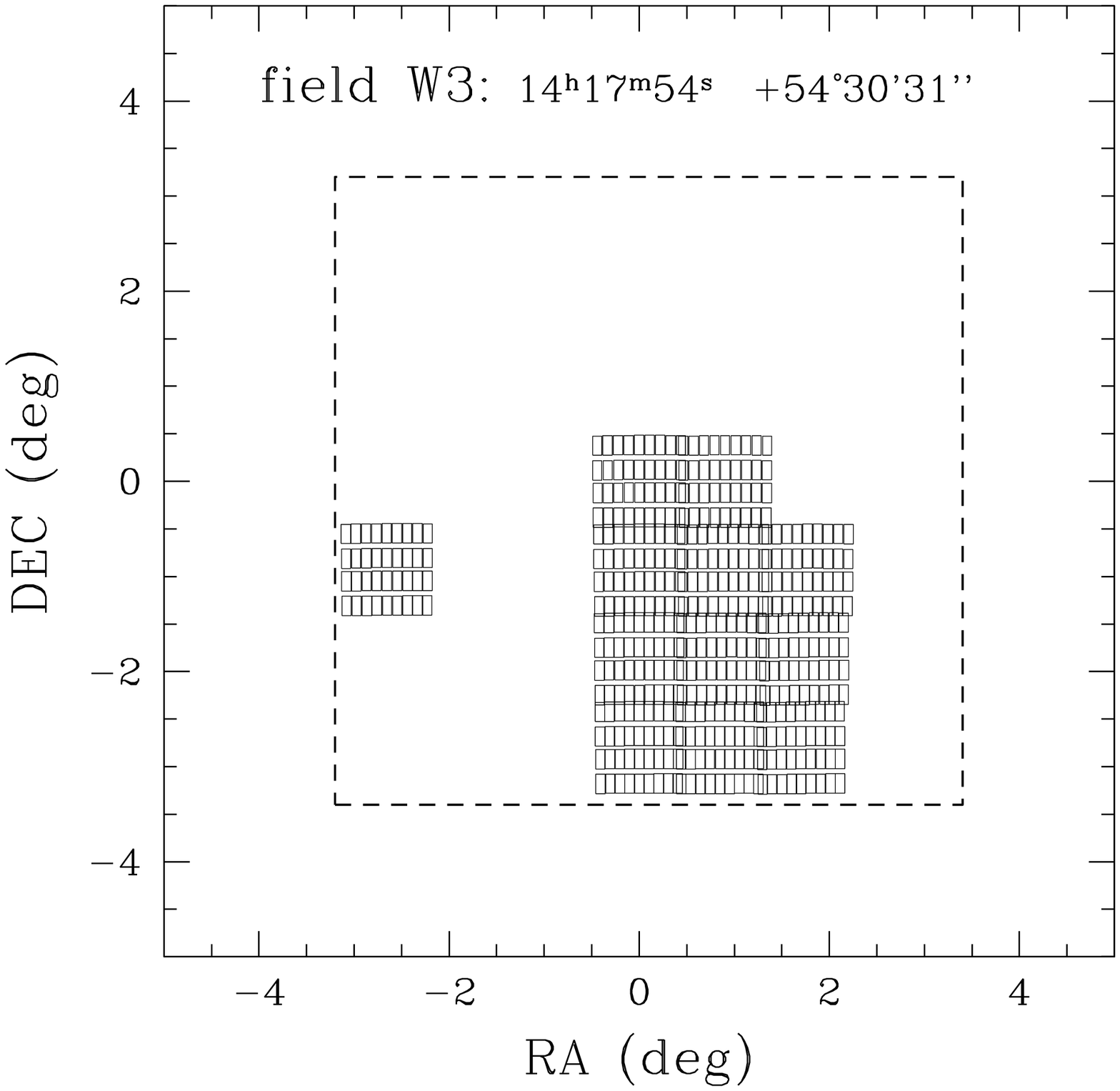}}
\figcaption{\footnotesize Layout of the observations of the W1 and W3
field. Each rectangle corresponds to one of the MegaCam chips. The
dashed box indicates the planned size of each of the two fields,
72 deg$^2$ for W1 and 49 deg$^2$ for W3. Currently 19 deg$^2$ of W1
and 12 deg$^2$ of W3 have been observed.
\label{layout}}
\end{center}
\end{figure*}

The reason for the recent popularity of cosmic shear is the fact that
the signal is a direct measure of the projected matter power spectrum
over a redshift range determined by the lensed sources and over scales
ranging from the linear to non-linear regime. This straightforward
interpretation of the signal is rather unique in the growing set of
tools available for cosmology, and is an important feature if we are
to determine the values of cosmological parameters with high
precision.  It enables us to measure important parameters such as the
matter density $\Omega_m$ and the amplitude of the power spectrum
$\sigma_8$, but also has the potential to constrain quintessence
models (Benabed \& van Waerbeke 2004; Tereno et al., in
preparation). The combination of cosmic shear with other well
understood probes, such as the cosmic microwave background (CMB) is
particularly powerful (e.g., Contaldi et al. 2003; Tereno et
al. 2005).

The measurement of the signal, however, is not without challenges,
which explains why cosmic shear has only recently appeared as a useful
tool in cosmology. First of all, the induced change in the shapes of
distant galaxies is small (less than 1\%), much smaller than the
intrinsic shapes of the sources themselves. As a result, large areas
of the sky need to be surveyed in order to reduce the statistical
errors. The first detections, reported only a few years
ago (Bacon, Refregier \& Ellis 2000; Kaiser, Wilson \& Luppino 2000;
van Waerbeke et al. 2000; Wittman et al. 2000), were based on
relatively small areas (few deg$^2$ at most). Recent developments in
the construction of wide field imaging cameras on 4m+ class telescopes
have made it possible to image much larger portions on the sky to
warrant accurate measurements of the lensing signal. For instance, van
Waerbeke, Mellier \& Hoekstra (2005) presented results based on $\sim
12$ deg$^2$ of deep imaging data from the VIRMOS-Descart Survey. Other
competitive results are based on shallower data, which cover a larger
area, such as the Red-sequence Cluster Survey (RCS; Hoekstra et al.,
2002c) based on 53 deg$^2$ and the 75 deg$^2$ CTIO Lensing Survey
(e.g., Jarvis et al. 2003; Jarvis et al. 2005).

The second challenge is the careful removal of observational
distortions, introduced by the telescope optics and atmosphere. Much
work has been devoted to deal with the point spread function (PSF) and
several correction schemes have been developed (e.g., Kaiser et
al. 1995; Bernstein \& Jarvis 2002; Refregier 2003) and tested (e.g.,
see Heymans et al. 2005 for the most up to date
discussion). Fortunately the separation of the signal into gradient
(`E'-mode) and curl (`B'-mode) components provides a non-trivial test
of the level of residual systematics, including the presence of
intrinsic aligments of galaxies or observational distortions in the
images. In addition, several other tests can be performed to test the
accuracy of the corrections (e.g., Hoekstra et al. 2002b; Bacon et
al. 2003; Heymans 2003; van Waerbeke et al. 2005; Heymans et
al. 2005). The recent results presented in van Waerbeke et al. (2005)
and Jarvis \& Jain (2005) have demonstrated that these systematics can
be accurately corrected for, resulting in measurements free of
`B'-modes.

The next step is to survey much larger areas on the sky to sufficient
depth in multiple filters, which enables us to probe the evolution of
the matter power spectrum.  The Canada-France-Hawaii-Telescope Legacy
Survey (CFHTLS) aims to image $\sim 170$ square degree in the
$i'$-filter down to $i'_{AB}=24.5$. This is comparable to the depth
reached in the VIRMOS-Descart survey, but an order of magnitude larger
in survey area, thus resulting in a significant reduction in
statistical errors.  In addition, the fields are observed in 4
additional filters as to ensure photometric redshift information for
the sources. This is an important part of the survey, as it enables
tests for intrinsic alignments and the study of the evolution of the
matter power spectrum. The latter significantly improves the
constraints on cosmological parameters, in particular on the equation
of state of the dark energy (Benabed \& van Waerbeke 2004)

In this paper we present the first results from the CFHTLS cosmic
shear program, based on 31 pointings, resulting in an effective area
of $\sim 22$ deg$^2$ of $i'$ data, already a significant increase
compared to previous work. The structure of the paper is as
follows. In \S2 we discuss the CFHTLS and the data used in this paper
as well as the data reduction process. The weak lensing analysis is
described in \S3. The resulting cosmic shear signal is presented in
\S4 and the implications for cosmological parameters are discussed in
\S5.

\setcounter{footnote}{0}

\section{Data}

The Canada-France-Hawaii Telescope Legacy Survey (CFHTLS)\footnote{
{\tt http://www.cfht.hawaii.edu/Science/CFHLS}} is a joint
Canadian-French program to make efficient use of Megaprime, the CFHT
wide field imager, and to address a number of fundamental problems in
astronomy. Megaprime, equipped with MegaCam, a 36 CCD mosaic camera
with a field of view of $\sim 1$ deg$^2$, enables us to obtain deep
images of large areas of the sky. The survey itself consists of three
independent parts, each with their own primary science driver. The survey
has been allocated more than 450 nights over a $\sim 5$ year
period. 

The results presented in this paper deal with data collected as part
of the wide synoptic survey, or `Wide Survey' for short. Once
completed, it will cover $\sim 170$ deg$^2$ in three patches of 49 to
72 square degrees through the whole filter set (u*, g', r', i', z')
down to i'=24.5. The survey allows the study of the large scale
structures and matter distribution in the universe through weak
lensing and the galaxy distribution. The former application is the
focus of this paper. We also use cosmic shear results based on
muli-color data from the Deep component of the survey, which are
described in detail in Semboloni et al. (2005).

For the analysis presented here we use 19 pointings in the i'-band
from the W1-field and 12 pointings from the W3-field, resulting in an
effective survey area 22 deg$^2$ after masking. The layout of the
current data with respect to the final survey is presented in
Figure~\ref{layout}. These data were obtained during the 2003B, 2004A
and 2004B observing semesters. In addition, 4 deg$^2$ of the W2 field
have been observed, but the area is not contiguous and we omit these
data from the current analysis.

\begin{figure*}
\begin{center}
\leavevmode
\hbox{%
\epsfxsize=8.5cm
\epsffile{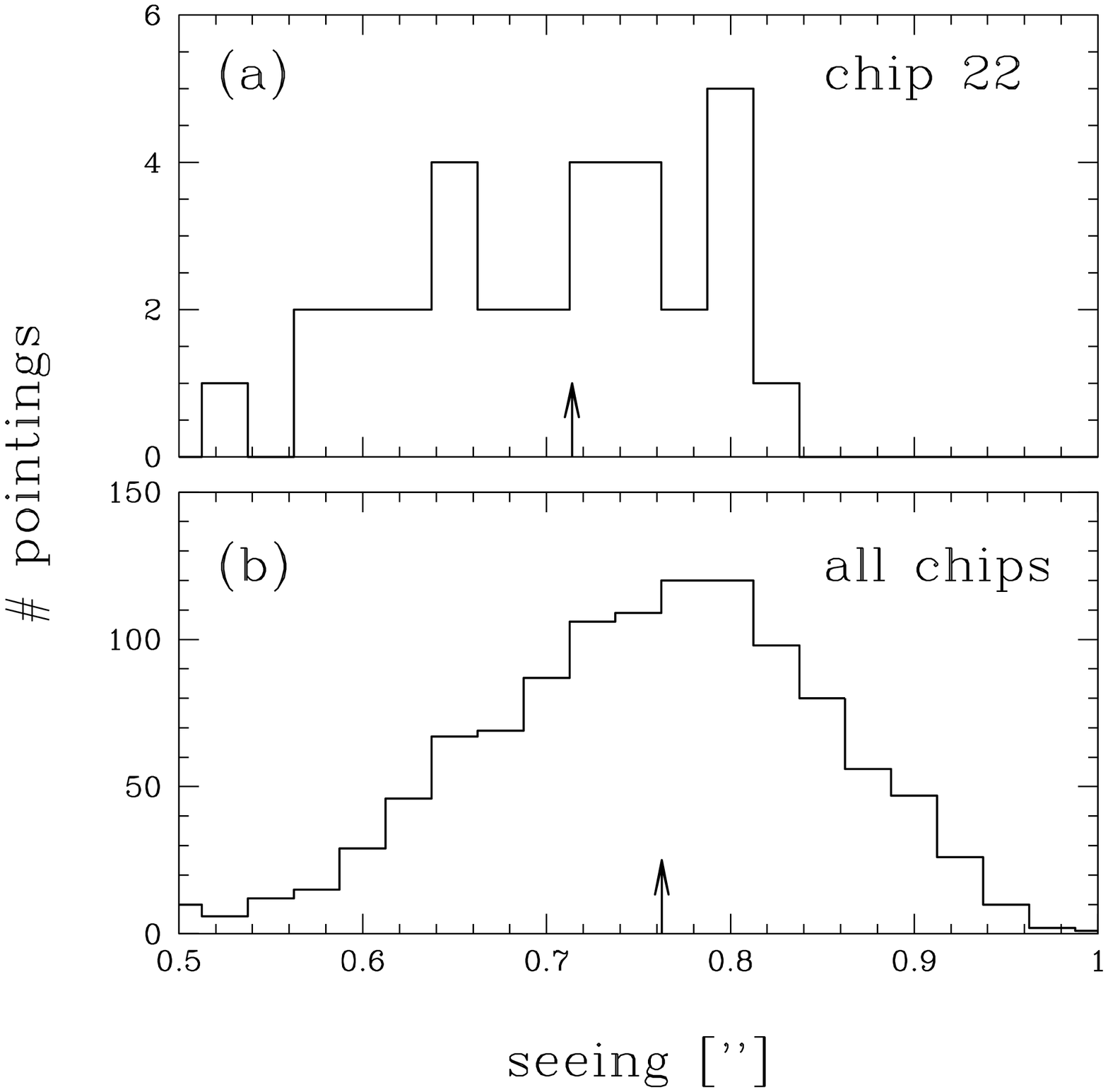}
\epsfxsize=8.5cm
\epsffile{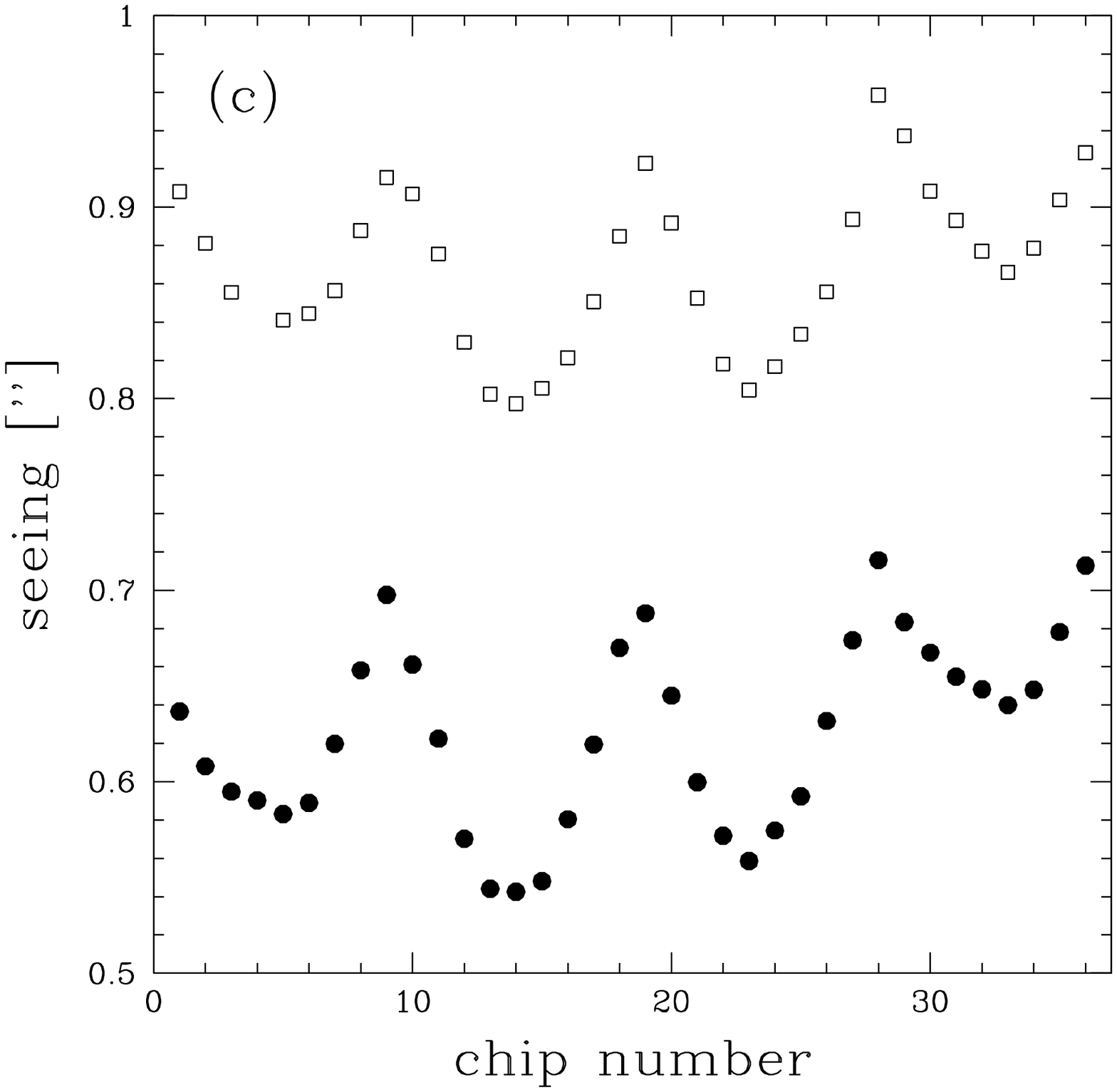}}
\figcaption{\footnotesize (a) FWHM of stars measured from the stacked images
of chip 22 for all pointings. This chip typically has the best image
quality (see panel~c). (b) FWHM measured for all chips, providing a
more accurate representation of the PSF size distribution. The arrows
indicate the median seeing for both distributions. (c) The PSF size
as a function of chip number for the best seeing image (filled circles)
and the worst seeing image (open squares). The camera consists of
4 rows of 9 chips each. The image quality is best in the center of the camera,
and degrades towards the edge as is indicated by these measurements.
\label{seeing}}
\end{center}
\end{figure*}

\subsection{Data Reduction}

Detrended data (de-biased and flatfielded) are provided by CFHT to the
community through the Canadian Astronomical Data Centre (CADC). The
detrending is done using the Elixir pipeline developed at CFHT. The
pipeline also provides photometric zeropoints and in most cases a
reasonable first order astrometric solution. 

The photometric calibrations are based on observations of standard
stars during the observing run. These zero-points are only valid under
photometric conditions. We therefore examine the magnitudes of a large
number of objects in the images, to check the stability of the
photometric zero-point and scale the images to the brightest
image. Most data were taken under photometric conditions, and the
corrections are found to be small.

The astrometric solutions provided by the Elixir pipeline are more
reliable for the more recent data, whereas a small fraction of chips
of the earliest observations require a significant revision of the
initial astrometry. The provided astrometric solution, however, is not
accurate enough for the purpose of our study and needs to be improved
upon. This process is too time consuming to be done manually, and is
done using an additional pipeline. The USNO-A2 catalog could be used
to refine the astrometry, but the number density of sources is often
too low to warrant stable results. Instead, retrieved a red image from
the second generation Digital Sky Survey (POSS~II) for each
pointing. These observations have small geometric distortions. The
astrometry of the POSS~II image is calibrated using the USNO-A2
catalog. SExtractor (Bertin \& Arnouts 1996 ) is used to generate a
catalog of sources with accurate astrometry, with a number density
significantly higher than the USNO-A2 catalog. In addition, the
POSS~II images have been taken more recently, thus reducing the
effects of proper motions of the stars.

This new astrometric catalog is matched to each of the MegaCam images.
The exposures have been taken with different offsets, in order to fill
the gaps between the chips. We combine the matched catalogs for each
exposure into a master catalog, which contains the average positions
of the matched objects. This master catalog is used to derive the
final second order astrometric solution for each chip. This procedure
ensures that in the overlapping area, the objects in each exposure are
accurately matched to the same position, which is crucial when
stacking the images for a weak lensing analysis: errors in the
astrometry lead to additional anisotropies in the images.

Tests show that the resulting astrometric solution for each pointing
is suffiently accurate to stack all data into a large image. However,
in this case different chips can contribute to the image at a given
position. If the PSF properties ``jump'' between chips, this gives
rise to a complicated PSF anisotropy pattern. Instead, we use only
those regions of the sky which in the final stacks were all observed
by the same chip. This avoids complicated behaviour of the PSF, at the
expense of approximately 20\% of the survey area. In the future we
plan to investigate in more detail to what extent we can deal with the
PSF on full mosaic images.

Before stacking the images, we identify stars on chip~22 (which has
the best image quality; see Figure~\ref{seeing}) and measure the
FWHM of the PSF and PSF anisotropy. If the seeing of some of the exposures is
significantly worse than the others, or if the images show extreme PSF
anisotropy, those images are discarded before stacking the images.  We
stack each chip separately, using the SWarp routine. The resulting
images are cropped such that the overlapping regions remain. These
images are used in the weak lensing analysis presented here. Typically
the stacked images consist of 7 images, each with an integration time
of 620s, but because of reasons described above, in a few cases we end
up with stacks of 6 images. The improvement in the PSF behavious
warrants the minor decrease in depth.

Figure~\ref{seeing} shows the seeing distribution (or PSF size) as
measured from the stacked images. Panel~a shows the distribution for
stars selected on chip~22, which typically has the best image
quality. For this particular chip we find a median seeing of
$0\farcs71$. Towards the edge of the field of view the image quality
degrades and a more accurate representation of the PSF size
distribution is presented in Figure~\ref{seeing}b, which shows a
histogram of the distribution based on all chips. In this case the
median FWHM is $0\farcs76$. We note that recent changes to Megaprime
have led to a significant improvement in image quality and a large
reduction in PSF anisotropy, thus reducing the level of systematics in
future cosmic shear measurements.

Nevertheless, the image quality in these early data is better than 1
arcsecond for all data used here. Figure~\ref{seeing}c shows the PSF
size as a function of chip number for the best seeing image (filled
circles) and worst seeing image (open squares). The camera consists of
four rows of nine chips, which results in the ``periodic'' changes in
the seeing. In both cases the PSF size is best in the middle of the
camera and increases towards the edges.

\section{Weak Lensing Analysis}

The weak lensing analysis presented here is based on the method
proposed by Kaiser et al. (1995) and Luppino \& Kaiser (1997) with a
number of modifications that are described in Hoekstra et al. (1998)
and Hoekstra et al. (2000). This particular method has been tested in
great detail (e.g., Hoekstra et al. 1998, 2002b; Bacon et al. 2001;
Erben et al. 2001; Heymans et al. 2005) and is widely used for cosmic
shear studies. Several promising alternative methods have been
developed recently (e.g, Bernstein \& Jarvis 2002; Refregier 2003) and
tests are underway to ensure accurate measurements of the lensing
signal.

Although the Kaiser et al. (1995) method is not the `ultimate'
technique to extract the lensing signal from the images, it has proven
to be one of the most accurate techniques currently available when
tested on simulated data as part of the Shear TEsting Programme (STEP;
Heymans et al. 2005). In this experiment, various weak lensing
pipelines were used to measure the shear in simulated images. The
simulated data consisted of a series with 5 different shears (constant
across the images) and 5 different (fairly realistic) PSFs, each
introducing different systematics.  Hence STEP provides a test of how
one can correct for both PSF anisotropy and the size of the PSF.
However, the simulations do not capture all details of real PSFs or
imperfections introduced by the stacking process. In this experiment,
the Hoekstra et al. (1998; 2000) implementation of the Kaiser et
al. (1995) approach was able to recover the lensing signal with an
accuracy better than 2\%. This accuracy is sufficient for published
cosmic shear results, and acceptable for the results presented here,
but we note that improvements are likely to be needed when analysing
the complete CFHTLS data set.

As mentioned earlier, we analyse the stacked images for each chip
individually. The first step in the analysis is the detection of
objects, for which we used the hierarchical peak-finding algorithm
from Kaiser et al. (1995). We select objects that are detected with a
significance greater than $5\sigma$ over the local sky. In addition to
the positions of the objects, the peak finder also provides fair
estimates of the object's size. We use this information to remove all
objects smaller than the PSF. Inspection shows that these are either
extremely faint objects, or, more relevant, spurious detections of
diffraction spikes, etc. The remaining objects are analysed in more
detail, which yields estimates for the size, apparent magnitude, and
shape parameters (polarization and polarizabilities). The apparent
magnitudes are corrected for galactic extinction using the results
from Schlegel et al. (1998).

The images were inspected by eye to mask out areas where the shape
measurements could be compromised. Potential sources can be cosmetic,
such as bleeding stars, halos, diffraction spikes, but also
astronomical, such as HII regions or spiral structure in resolved
galaxies. We separate the stars and galaxies on the basis of their
half-light radii. The galaxy shapes are corrected for observational
distortions as described below. The final, corrected, catalogs for
each chip are combined into a large catalog for each field.

For the weak lensing analysis we use only galaxies brighter than
$i'_{AB}=24.5$, which leaves a sample of $9.7\times 10^5$ galaxies in
the W1 field and $6.5\times 10^5$ galaxies in the W3 field. The
surveyed area is $\sim 22$ deg$^2$, resulting in an average galaxy
number density of $\sim 20$ galaxies arcmin$^{-2}$. However, the shape
measurement errors are larger for faint, small galaxies. Therefore
these galaxies should be given less weight for an optimal estimate of
the lensing signal (Hoekstra et al. 2000).

Instead it is more convenient to define an effective number density of
galaxies $n_{\rm eff}$, which is related directly to the measurement
error in the shear for an area of 1 arcmin$^2$, which is the relevant
quantity.  Hubble Space Telescope observations indicate that for well
resolved galaxies the intrinsic shapes of galaxies results in a
dispersion $\langle\gamma^2\rangle^{1/2}\approx 0.3$ (Hoekstra et
al. 2000).  If all galaxies were measured `perfectly', the error in
the shear measurement would be given by
$\sigma_\gamma=0.3/\sqrt{n_{\rm eff}}$. For the CFHTLS data this yields
an effective number density of $\sim 12$ galaxies arcmin$^{-2}$.

\begin{figure}
\begin{center}
\leavevmode
\hbox{%
\epsfxsize=8.5cm
\epsffile[20 180 480 633]{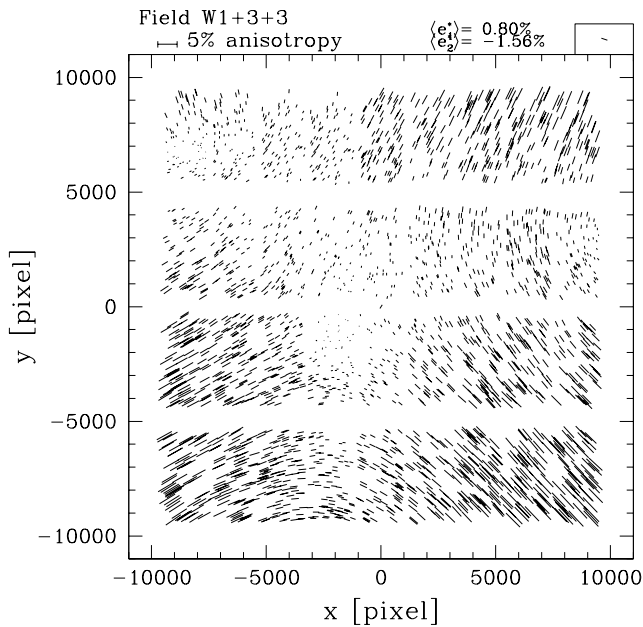}}
\figcaption{\footnotesize A typical example of the PSF anisotropy as a
function of position for a stacked Megacam image. The sticks indicate
the direction of the major axis of the PSF and the length is
proportional to the observed ellipticity of the PSF. In order to show
the higher order spatial dependence of the anisotropy, we have
subtracted the average ellipticity. The direction of the average PSF
anisotropy is indicated in the top right box and the amplitude is
indicated as well.  Although the PSF anisotropy is determined for
individual chips, the figure clearly shows a large-scale coherent
pattern.
\label{psfan}}
\end{center}
\end{figure}

\subsection{Correction for PSF Effects}

The observed shapes of the galaxies cannot be used to measure the
lensing signal, because observational distortions have significantly
altered their shapes in a systematic fashion: PSF anisotropy
introduces coherent aligments in the galaxy shapes and the seeing
circularizes the images. The key to an accurate measurement of the
weak lensing signal lies in the adequate correction for these
systematic effects. As mentioned above, our pipeline has been tested
extensively (e.g., Hoekstra et al. 2002b; Heymans et al. 2005) and has
been shown to be able to recover the weak lensing shear with an
accuracy of $\sim 2\%$.

The first step in this procedure is the identification of a sample of
moderately bright stars which can be used to quantify the properties
of the PSF (anisotropy and `size'). The pattern of PSF anisotropy
changes from observation to observation, although we found that the
Megacam PSF is relatively stable, in particular when compared to the
previous CFH12k camera (e.g., Hoekstra 2004). The PSF anisotropy also
varies across the field of view and this spatial variation is captured
by fitting a second order polynomial to the shape parameters of the
stars for each chip.

We found that a second order polynomial provided an excellent fit to
the data. Note that the choice of model depends on the properties of
the camera. For instance van Waerbeke et al. (2005) used rational
functions (as suggested by Hoekstra (2004)) to describe the pattern of
the CFH12k camera. We intend to improve our characterization following
the procedure developed by Jarvis \& Jain (2005), which combines
information from a series of exposures, effectively resulting in a
denser sampling of the PSF variation.

A typical example of the PSF anisotropy as a function of position
across the Megacam mosaic is presented in Figure~\ref{psfan}. To show
the spatial variation in more detail, we have substracted the mean
anisotropy across the field. The resulting pattern is coherent across
the field, even though the fits were obtained from individual chips.

Having quantified the PSF anisotropy and its spatial variation, we can
undo its effect following Kaiser et al. (1995) and Hoekstra et al.
(1998). The same stars used to study the PSF anisotropy are also used
to correct for the dilluting effect of seeing, as described in Luppino
\& Kaiser (1997) and Hoekstra et al. (1998).

\begin{figure*}
\begin{center}
\leavevmode
\hbox{%
\epsfxsize=8.5cm
\epsffile{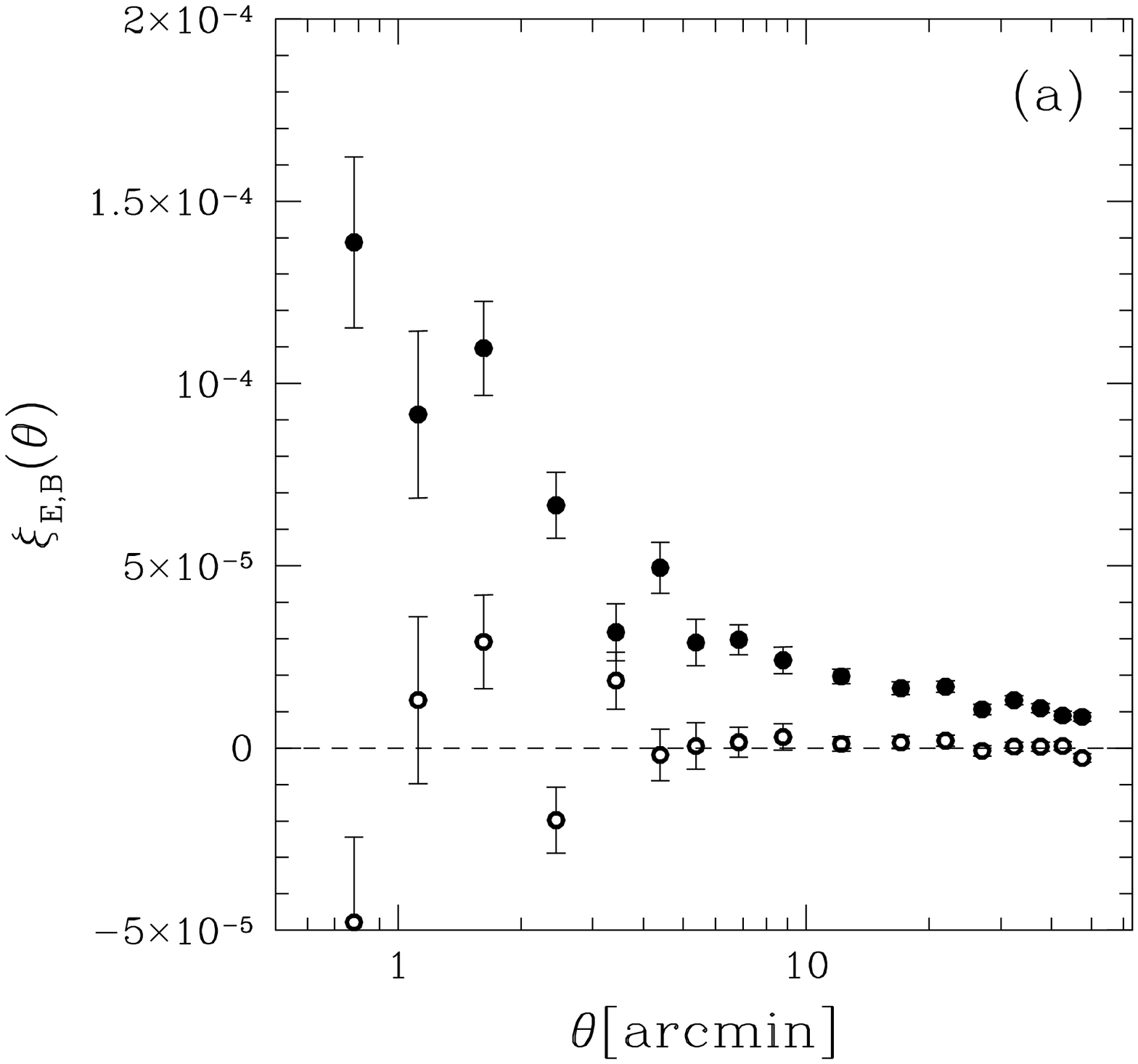}
\epsfxsize=8.5cm
\epsffile{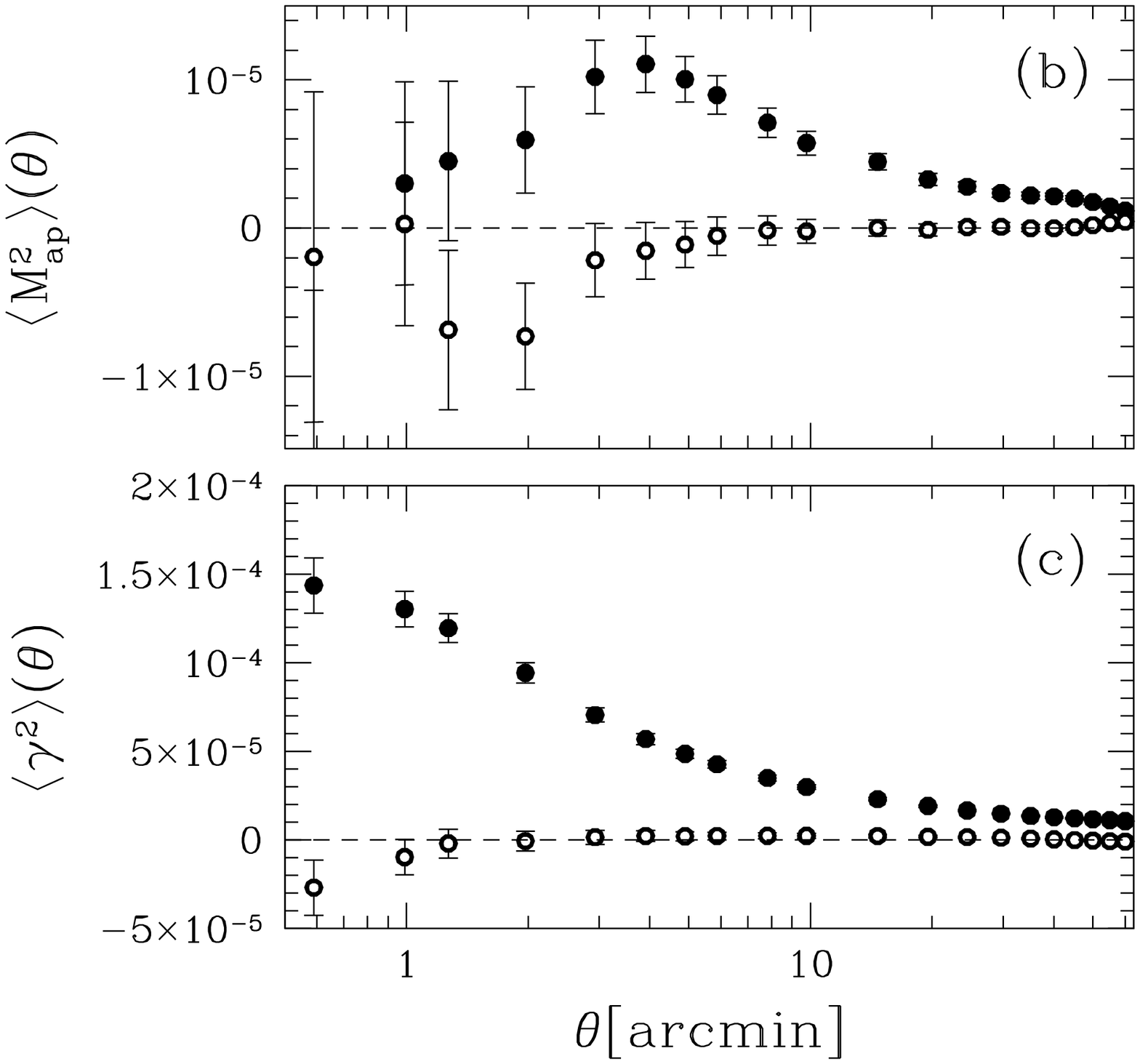}}
\figcaption{\footnotesize Panel a) `E' and `B'-mode shear correlation
functions (filled and open points, respectively) measured from $\sim
22$ deg$^2$ of $i'$ data from the CFHT Legacy Survey. The error bars
only indicate the statistical errors determined for the combined
signal from the data from W1 and W3 fields.  Panel b) `E' and `B'-mode
measurements for the aperture mass statistic. Note that the `B'-mode
is consistent with zero on all scales. Panel c) as before but now for
the top-hat variance.
\label{twopoint}}
\end{center}
\end{figure*}

\section{Measurements}

To quantify the lensing signal, we measure the ellipticity (or shear)
correlation functions from the galaxy shape catalogs. These
correlation functions, in turn, can be related to the various
two-point statistics that are commonly used in the literature. The use
of the ellipticity correlation functions also allow for the separation
of the signal into `E'-mode (gradient) and `B'-mode (curl) components.
Gravitational lensing arises from a gravitational potential and it is
therefore expected that the lensing signal is curl-free. The amplitude
of the `B'-mode then provides a measure of residual systematics.

In \S5 we briefly discuss how these observable two-point statistics
relate to the matter power spectrum and cosmology. For more detailed
discussions we refer to Schneider et al. (1998) and Bartelmann \&
Schneider (2001).

\noindent The two ellipticity correlation functions that are measured are

\begin{equation}
\xi_{\rm tt}(\theta)=\frac{\sum_{i,j}^{N_s} w_i w_j
\gamma_{{\rm t},i}({{\bf x}_i}) \cdot \gamma_{{\rm t},j}({{\bf x}_j})}
{\sum_{i,j}^{N_s} w_i w_j},
\end{equation}

\noindent and

\begin{equation}
\xi_{\rm rr}(\theta)=\frac{\sum_{i,j}^{N_s} w_i w_j
\gamma_{{\rm r},i}({{\bf x}_i}) \cdot \gamma_{{\rm r},j}({{\bf x}_j})}
{\sum_{i,j}^{N_s} w_i w_j},
\end{equation}

\noindent where $\theta=|{\bf x}_i-{\bf x}_j|$. $\gamma_{\rm t}$ and
$\gamma_{\rm r}$ are the tangential and 45 degree rotated shear in the
frame defined by the line connecting the pair of galaxies. The weights
$w_i$ are proportional to the inverse square of the uncertainty in the
shear (Hoekstra et al. 2000). For the following, it is more useful to
consider

\begin{equation}
\xi_\pm(\theta)=\xi_{\rm tt}(\theta)\pm \xi_{\rm rr}(\theta),
\end{equation}

\noindent i.e., the sum and the difference of the two observed
correlation functions. As shown by Crittenden et al. (2002), one can
derive `E' and `B'-mode correlation functions by integrating
$\xi_+(\theta)$ and $\xi_-(\theta)$ with an appropriate window
function

\begin{equation}
\xi^E(\theta)=\frac{\xi_+(\theta)+\xi'(\theta)}{2}~{\rm and~}
\xi^B(\theta)=\frac{\xi_+(\theta)-\xi'(\theta)}{2},
\end{equation}

\noindent where

\begin{equation}
\xi'(\theta)=\xi_-(\theta)+4\int_\theta^\infty \frac{d\vt}{\vt}\xi_-(\vt)
-12\theta^2\int_\theta^\infty \frac{d\vt}{\vt^3}\xi_-(\vt).
\end{equation}

We also present results for other frequently used two-point
statistics, namely the aperture mass variance $\langle M_{\rm
ap}^2\rangle(\theta)$ and the top-hat smoothed variance
$\langle\gamma^2\rangle(\theta)$.

Of these statistics, the aperture mass is of some particular interest,
because only for this case are the `E' and `B'-modes uniquely defined,
whereas the decompositions of the shear correlation function and the
top-hat variance in E and B modes are defined up to a constant
(Crittenden et al. 2002; Pen et al. 2002). The aperture mass is
defined as

\begin{equation}
M_{\rm ap}(\theta)=\int d^2\vt U(\vt) \kappa(\vt),
\end{equation}

\noindent where $\kappa$ is the dimensionless surface density or
convergence. Provided $U(\vt)$ is a compensated filter (i.e., a filter
such that a constant surface density within the aperture yields
$M_{\rm ap}=0$), the aperture mass can be expressed in term of the
observable tangential shear $\gamma_{\rm t}$ using a different filter
function $Q(\vt)$ (which is a function of $U(\vt)$),

\begin{equation}
M_{\rm ap}(\theta)=\int_0^\theta d^2\phi Q(\vt)\gamma_{\rm t}(\vt).
\end{equation}

\noindent We use the filter function suggested by Schneider et al. (1998)

\begin{equation}
U(\theta)=\frac{9}{\pi\theta^2}
\left(1-\frac{\vt^2}{\theta^2}\right)
\left(\frac{1}{3}-\frac{\vt^2}{\theta^2}\right),
\end{equation}

\noindent for $\theta\le\vt$, and 0 elsewhere. The corresponding $Q(\theta)$
is given by

\begin{equation}
Q(\theta)=\frac{6}{\pi\theta^2}\left(\frac{\vt^2}{\theta^2}\right)
\left(1-\frac{\vt^2}{\theta^2}\right),
\end{equation}

\noindent for $\theta\le\vt$, and 0 elsewhere. The `E' and
`B'-mode aperture masses are computed from the ellipticity
correlation functions using

\begin{equation}
\langle M_{\rm ap}^2\rangle(\theta)=\int d\vt~\vt \left[{\cal W}(\vt)\xi_+(\vt)+
\tilde{\cal W}(\vt)\xi_-(\vt)\right],
\end{equation}

\noindent and

\begin{equation}
\langle M_\perp^2\rangle(\theta)=\int d\vt~\vt \left[{\cal W}(\vt)\xi_+(\vt)-
\tilde{\cal W}(\vt)\xi_-(\vt)\right],
\end{equation}

\noindent where ${\cal W}(\vt)$, and $\tilde{\cal W}(\vt)$ are given
in Crittenden et al. (2002). Useful analytic expressions were derived
by Schneider et al. (2002a). Both ${\cal W}(\vt)$, and $\tilde{\cal
W}(\vt)$ vanish for $\vt>2\theta$, so that $\langle M_{\rm
ap}^2\rangle$ can be obtained directly from the observable ellipticity
correlation functions over a finite interval. Similarly, the top-hat
variance can be expressed in terms of the ellipticity correlation
functions, but with different filter functions, as described in
Schneider et al. (2002a).

\subsection{Lensing signal}

The results for the various two-point statistics are presented in
Figure~\ref{twopoint}. The filled points in Figure~\ref{twopoint}a
indicate the observed `E'-mode shear correlation function as a
function of angular scale. The error bars indicate only the
statistical uncertainty in the measured signal, and ignores any
contribution from cosmic variance which is present in the `E'-mode
signal. However, when estimating cosmological parameters, as discussed
in \S5, we include estimates for cosmic variance. The open points
correspond to the `B'-mode. The signal presented in
Figure~\ref{twopoint} is based on the 22 deg$^2$ of data for the W1
and W3 fields.

As mentioned previously, in order to separate the signal into `E' and
`B'-modes, we have to define a zero-point for the `B'-mode for the
shear correlation function and top-hat variance. Based on the absence
of `B'-modes in the observed aperture mass variance, shown in
Figure~\ref{twopoint}b, the correlated function in panel~a has been
scaled such that the mean `B'-mode is zero on scales larger than 10
arcminutes. 

Figure~\ref{twopoint}c shows the top-hat variance $\langle
\gamma^2\rangle$ as a function of scale.  All three statistics have
been used in the past to estimate cosmological parameters, and each
has its own distinct advantage and disadvantage. For instance the
aperture mass provides an absolute calibration of the `B'-mode, but is
most sensitive to the matter power spectrum on small scales. To probe
the largest scales available in the surveyed area, we chose to use
the top-hat variance for our estimates of cosmological parameters,
presented in \S5.

\begin{figure}
\begin{center}
\leavevmode
\hbox{%
\epsfxsize=8.5cm
\epsffile{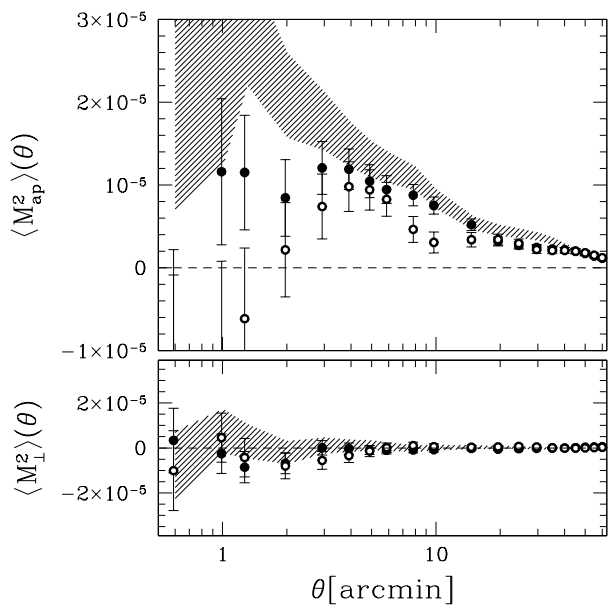}}
\figcaption{\label{mapcomp} {\it top panel:} Aperture mass variance
determined from W1 (solid points) and W3 (open points) separately.
For reference, the shaded region indicates the $1\sigma$ area
around the measurement from the VIRMOS-Descart survey (van Waerbeke
et al. 2005), which is of similar depth. The W1 measurements agree
well with the latter, but the W3 signal is lower than expected.
On large scales (beyond 10 arcminutes) the agreement between all
three measurements is excellent. Note, however, that
the error bars do not include cosmic variance. {\it bottom panel:} Aperture
mass `B'-mode for all three measurements, which are all consistent
with no signal.}
\end{center}
\end{figure}

The results presented in Figure~\ref{twopoint} combine the
measurements of the W1 and W3 fields. However, it is useful to compare
the signals obtained from the individual fields as well, to check for
consistency. The aperture mass variances for the two fields are shown
in the upper panel of Figure~\ref{mapcomp}. The solid points
correspond to W1, whereas the open points are for W3. For reference we
also show the results from van Waerbeke et al. (2005), indicated by
the shaded region. The latter uses a similar range in apparent
magnitude,although the filter is different ($I_C$ vs. $i'$ used here).
Hence the amplitude should be comparable to our measurements.  On
scales beyond 10 arcminutes all three measurements are in good
agreement, but on small scales, the results from W3 appear lower than
expected. The origin of this discrepancy is not clear, and increasing
the surveyed area for this field might provide a better understanding.
The lower panel of Figure~\ref{mapcomp} shows the corresponding
`B'-modes, which are all consistent with zero on all scales.

\begin{figure}
\begin{center}
\leavevmode
\hbox{%
\epsfxsize=8.5cm
\epsffile{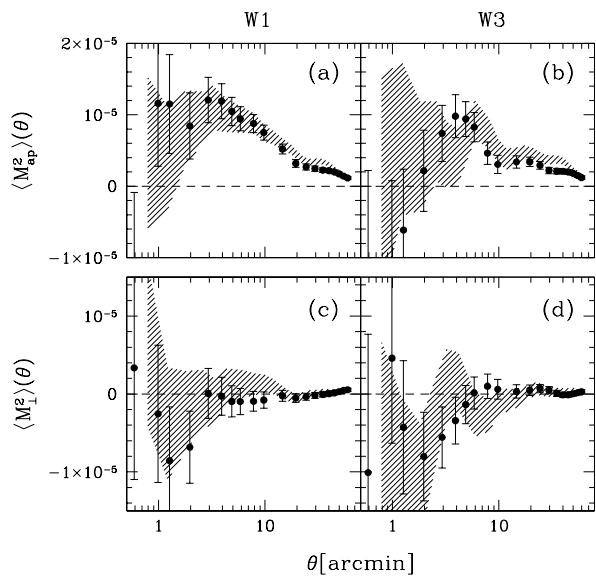}}
\figcaption{\label{paris} {\it panels a and b}: The aperture mass variance
from the W1 and W3 fields, respectively. The points correspond to the
results presented here. The shaded regions indicate the $1\sigma$ area
around the measurements from the Paris pipeline (Fu et al., in preparation).
{\it panels c and d}: The aperture mass `B'-mode signal for the W1 and
W3 fields. The results from the two pipelines agree very well. Note, that
the Paris results are based on a smaller survey area than the one presented
in this paper.}
\end{center}
\end{figure}

This paper present the first CFHTLS results based on one of two
reduction and analysis pipelines. An independent analysis was
performed by the Paris group and for reference we also show a
comparison with their preliminary results in Figure~\ref{paris}.  A
detailed discussion of this alternative analysis will be presented in
Fu et al. (in preparation). The top panels in Figure~\ref{paris} show
the aperture mass variance for the W1 and W3 fields. The points
correspond to the results from the analysis presented in this paper.
The shaded regions indicate the $1\sigma$ area around the measurements
from the Paris pipeline (Fu et al., in preparation). The lower panels
show the corresponding measurements of the `B'-mode signals. The
results from the two pipelines agree very well. We note, however, that
the Paris results are based on conservatively masked images, resulting
in a smaller effective survey area than the one presented in this
paper.

\subsection{Tests for Residual Systematics}

The absence of a significant `B'-mode is very encouraging. However, by
itself it cannot guarantee that the results are free of systematics
(the presence of a `B'-mode provides strong evidence for residual
systematics, but the converse is not true). Fortunately several other
tests can be performed as well.  These tests essentially provide
additional checks on the correction for PSF anisotropy, and we present
results in this section.

If the correction for the size of the PSF (i.e., seeing) is imperfect,
this can result in calibration biases (e.g., Hirata \& Seljak 2003).
Unfortunately, there is no good test of this correction, based on the
data alone. Instead, one needs to rely on comparison with simulated
data sets, such as STEP (Heymans et al. 2005). As mentioned above, the
results of the latter experiment suggests that we can recover the
shear to better than $2\%$. We will assume a systematic error of this
magnitude when estimating cosmological parameters.

Depending on the set of cosmological parameters one is interested in,
one can marginalize over the uncertainty in this correction, with
little loss in accuracy (Ishak et al. 2004). This approach, however,
may fail when studying the effect of massive neutrinos (e.g.,
Abazajian \& Dodelson 2003). However, such an approach does not take
into account the variable bias that might arise from pointing to
pointing seeing variations. Fortunately, as shown in Vale et
al. (2004), the signal is not affected significantly by small
variations in calibration errors, thus reducing the complexity of the
problem for current data sets.

\begin{figure}
\begin{center}
\leavevmode
\hbox{%
\epsfxsize=8.0cm
\epsffile{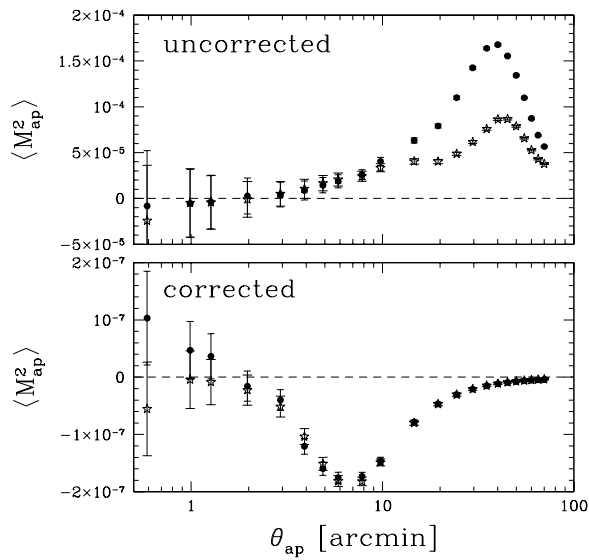}}
\figcaption{\label{psf_map} {\it top panel}: The aperture mass variance
computed from the uncorrected shapes of the stars. The `E'-mode (solid
points) and `B'-modes (stars) are very similar on scales $<10$ arcminutes,
but on larger scales the `E'-mode is larger. {\it bottom panel}: The
aperture mass using the residual shapes of stars after correcting for PSF
anisotropy. The `E' and `B' modes are indistinguishable on all scales.
The results suggest that the adopted PSF model is inadequate on scales
between 4 and 20 arcminutes. Note, however, that the lensing signal
on these scales is $\sim 30$ times larger than the residuals presented
here.}
\end{center}
\end{figure}

As was shown in Hoekstra (2004) and van Waerbeke et al. (2005) the PSF
anisotropy leads to a systematic signal which can have different `E'
and `B'-modes. For instance, the systematic `B'-mode was $\sim 3$
times lower than the systematic `E'-mode. Furthermore the source of an
imperfect correction for PSF anisotropy can be two-fold. First of all,
the correction method itself might introduce errors. This has been
studied in great detail (e.g. Hoekstra et al. 1998; Heymans et
al. 2005) demonstrating that the correction scheme used here, is
sufficiently accurate. Secondly, as pointed out in Hoekstra (2004), an
incorrect model for the spatial variation of the PSF anisotropy can
lead to considerable errors.  

Recently, Jarvis \& Jain (2005) have presented an interesting way to
improve the PSF model using the large number of exposures taken in a
weak lensing survey. They analyse the PSF variation using a principal
component analysis to identify the dominant PSF anisotropy
patterns. Under the assumption that only a few parameters are needed
to describe most of the variation, this approach allows for a much
more detailed modelling of the PSF. We plan to implement this method
for future analyses of the CFHTLS data.

\begin{figure}
\begin{center}
\leavevmode
\hbox{%
\epsfxsize=8.5cm
\epsffile{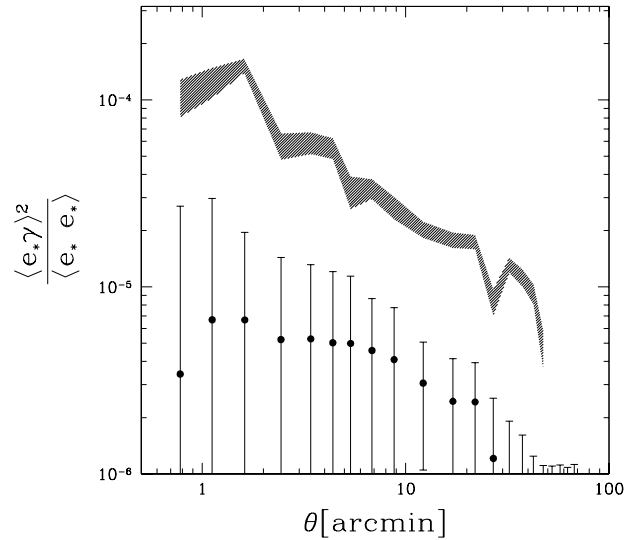}}
\figcaption{\label{xi_sys} The points with error bars show the
residual systematics correlation function $\xi_{\rm sys}$ as defined
in the text.  The indicated error bars are those of the shear
correlation function, and are displayed to show the level of
systematics with respect to the $1\sigma$ statistical error. The
shaded region corresponds to the $1\sigma$ region around the observed
shear correlation itself, which is an order of magnitude larger than
$\xi_{\rm sys}$}
\end{center}
\end{figure}

We can test both of these potential points of failure. The accuracy of
the model describing the spatial variation of the PSF anisotropy can
be examined by correcting the measured shapes of the stars and
computing the aperture mass variance of the residuals.  In this case,
any residual signal is caused by imperfections of the model (Hoekstra
2004). Figure~\ref{psf_map} shows the results of this test. For
reference, the upper panel shows the aperture mass variance of the
stars before correction for PSF anisotropy. The `E'-mode (solid
points) and `B'-modes (stars) are very similar on scales $<10$
arcminutes, but on larger scales the `E'-mode is significantly larger.
Given that this signal is larger than the cosmic shear signal on
scales larger than 10 arcminutes, this figure clearly demonstrates the
importance of a careful correction for PSF anisotropy.

The bottom panel of Figure~\ref{psf_map} shows the aperture mass
variance using the residual shapes of stars after they have been
corrected for PSF anisotropy. The `E' and `B' modes are
indistinguishable on all scales. The results clearly demonstrate that
the adopted PSF model is inaccurate on scales between 4 and 20
arcminutes. Note, however, that the level of the residuals is about a
factor of 30 smaller than the cosmic shear signal. Although there is
clearly room for improvement regarding the modelling of the spatial
variation of the PSF, the results presented in Figure~\ref{psf_map}
suggest that this is not a dominant source of error in our current
analysis.

The next step is to examine how well the shapes of the background
galaxies have been corrected. The amount of residual systematics left
in the weak lensing signal due to imperfect PSF correction can be
estimated from the correlation between the uncorrected stars and the
corrected galaxies. Bacon et al. (2003) and Heymans et al. (2003)
defined a useful estimator

\begin{equation}
\xi_{\rm sys}=\frac{\langle e^\star\gamma\rangle^2}{\langle e^\star e^\star\rangle},
\end{equation}

\noindent where $e^\star$ is the ellipticity of the stars before PSF
correction and $\gamma$ is the shear estimate of the galaxies. The
estimator is conveniently normalized by the star ellipticity
auto-correlation function, which allows for a direct comparison to the
lensing signal $\langle\gamma(r)\gamma(\theta+r)\rangle$. Note that
this estimator is sensitive to imperfections in the model for PSF
anisotropy {\it and} imperfections in the correction scheme itself.

The points in Figure~\ref{xi_sys} correspond to the resulting value
for $\xi_{\rm sys}$ as a function of scale. The indicated error bars
are those of the shear correlation function, and are displayed to show
the level of systematics with respect to the $1\sigma$ statistical
error in the lensing signal.  The shaded region corresponds to the
$1\sigma$ region around the observed shear correlation itself, which
is an order of magnitude larger than $\xi_{\rm sys}$. These results
demonstrate that the signal presented in Figure~\ref{twopoint} is not
significantly affected by systematics and can be used to provide
reliable constraints on cosmological parameters. Note, however, as
the CFHTLS progresses the level of residual systematics needs to be
reduced.

\section{Cosmological parameters}

The observed two-point statistics can be related to the matter power
spectrum $P_\delta(k)$, which depends on a range of cosmological
parameters. For the study presented here, the most relevant parameters
are the matter density $\Omega_m$ and the normalisation $\sigma_8$.
In the case of cosmic shear, some of the dependence on cosmology also
enters through the angular diameter distances to the sources.

\subsection{Method}

As discussed above, in order to probe the largest scales available in
the surveyed area, we choose to use the top-hat variance
$\langle\gamma^2\rangle$, which can be expressed in terms of the power
spectrum through

\begin{equation}
\langle\gamma^2\rangle(\theta)=2\pi\int_0^\infty
dl~l~P_\kappa(l)\left[\frac{J_1(l\theta)}{\pi l \theta}\right]^2,
\end{equation}

\noindent where $\theta$ is the radius of the aperture used to compute
the variance, and $J_1$ is the first Bessel function of the first
kind. $P_\kappa(l)$ is not the power spectrum itself, but the
convergence power spectrum, defined as 

\begin{equation} 
P_\kappa(l)=\frac{9 H_0^4 \Omega_m^2}{4 c^4}
\int\limits_0^{w_H}dw \left(\frac{\bar W(w)}{a(w)}\right)^2
P_\delta\left(\frac{l}{f_K(w)};w\right),
\end{equation}

\noindent where $w$ is the radial (comoving) coordinate, $w_H$
corresponds to the horizon, $a(w)$ the cosmic scale factor, and
$f_K(w)$ the comoving angular diameter distance. $\bar W(w)$ is the
source-averaged ratio of angular diameter distances $D_{ls}/D_{s}$ for
a redshift distribution of sources $n(w)$:

\begin{equation}
\bar W(w)=\int_w^{w_H} dw' n(w')\frac{f_K(w'-w)}{f_K(w')}.
\end{equation}

Hence, it is important to know the redshift distribution of the
sources, in order to relate the observed lensing signal to
$P_\kappa(l)$ and consequently, cosmological parameters.

For our selection of background galaxies $(21.5<i'<24.5)$, the
redshift distribution should be very similar to the one used by van
Waerbeke et al. (2005) in their analysis of the VIRMOS-Descart data.
Currently, only $i'$ images have been processed, but as the survey
progresses photometric redshifts for the sources will be determined,
which greatly enhances our ability to constrain the cosmology, and to
minimize the uncertainty in the source redshift distribution.

Unfortunately, the current observational constraints on the redshift
distribution are still limited, although a number of redshift surveys
are targeting the faint galaxies and redshift range used for weak
lensing (e.g., DEEP2: Davis et al. 2003; VVDS: Le F{\`e}vre et al.,
2004).  For the analysis here, the best information on the source
redshifts comes from photometric redshift studies of the Hubble Deep
Fields (Fern{\'a}ndez-Soto et al., 1999). It is convenient to
parametrize the redshift distribution using

\begin{equation}
n(z)=\frac{\beta}{z_s \Gamma\left(\frac{1+\alpha}{\beta}\right)}
\left(\frac{z}{z_s}\right)^\alpha \exp\left[-\left(\frac{z}{z_s}\right)^\beta
\right].
\end{equation}

\begin{figure}
\begin{center}
\leavevmode
\hbox{%
\epsfxsize=8.0cm
\epsffile{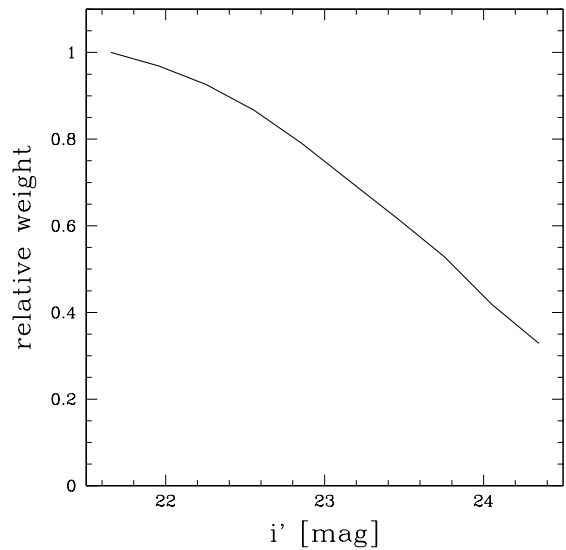}}
\figcaption{\label{weight} Average weight per galaxy as a function of
apparent $i'$-magnitude. Note that the weights are not normalised}
\end{center}
\end{figure}

The best fit parameters are obtained from a least squares fit to the
photometric redshift distribution, taking into account the adopted
weights for the shape measurements of the sources (see the discussion
in \S4). The weight is predominantly a function of apparent magnitude
and the average value is shown in Figure~\ref{weight}. As fainter
galaxies are, on average, at higher redshift, such a weighting scheme
will modify the true source redshift distribution into an `effective'
one. To compute the best fit parameters, we use the Poisson errors in
the counts as a function of redshift, thus ignoring field-to-field
variation in the HDF redshift distributions.

\begin{figure}
\begin{center}
\leavevmode
\hbox{%
\epsfxsize=8.5cm
\epsffile{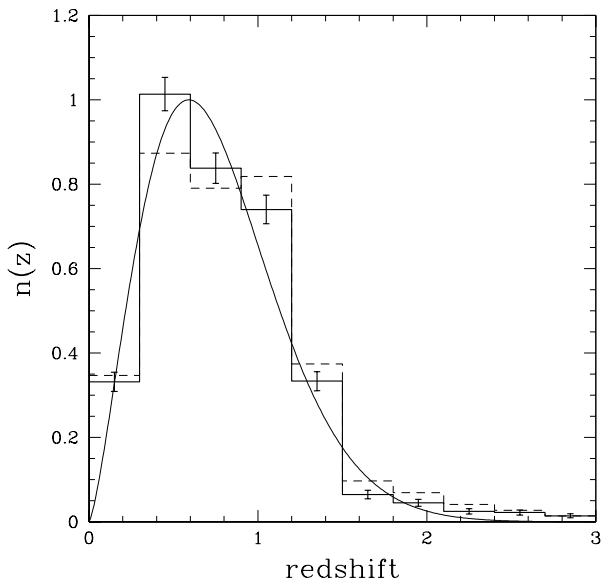}}
\figcaption{\label{zdist} Effective redshift distribution of galaxies
with $21.5< I_{AB} < 24.5$ determined from the Hubble Deep Field North
and South. The error bars are the Poisson errors from the finite
number of galaxies in each bin (note that these do not include 
cosmic variance). The smooth solid curve represents the
best fit model with a mean source redshift of $\langle
z\rangle=0.81$. The dashed histogram corresponds to the redshift
distribution when the adopted weighting scheme is not taken into
account.}
\end{center}
\end{figure}

The solid histogram in Figure~\ref{zdist} shows the effective source
redshift distribution. Comparison with the dashed histogram (which
corresponds to the unweighted case) shows that the weighting scheme
slightly lowers the mean source redshift. The smooth curve corresponds
to the best fit redshift distribution, which has parameters
$\alpha=1.35$, $\beta=1.654$ and $z_s=0.668$. This corresponds to a
mean source redshift of $\langle z\rangle=0.81$. To quantify the
uncertainties in the source redshift distribution, we keep $\alpha$
and $\beta$ fixed, but vary $z_s$ to identify the 68\% and 95\%
confidence intervals. This yields $z_s\in[0.632,0.703]$ with
68\% confidence and $z_s\in[0.613,0.721]$ with 95\% confidence.
We marginalise over the latter interval when estimating the cosmological
parameters.

The scales probed by the measurement are affected significantly by the
non-linear growth of structure. As shown by Jain \& Seljak (1997) and
Schneider et al. (1998), we cannot use the linear power spectrum, but
it is necessary to use the non-linear power spectrum. Two different
approaches to calculate the non-linear power spectrum have been
proposed and we present results for both. The first is based on the
scaling formula suggested by Hamilton et al. (1991), which was
extended to a wider range of cosmologies by Peacock \& Dodds (1996).
Peacock \& Dodds (1996) provide a prescription to compute the power
spectrum, using a fitting formula which is calibrated using numerical
simulations. More recently, Smith et al. (2003) suggested an approach
based on a halo model approach to better capture the breakdown of the
stable clustering assumption in the Peacock \& Dodds (1996)
prescription.  We note that both prescriptions are based on relatively
small numbers of numerical simulations and that their accuracy is
limited. Comparison with recent numerical simulations suggest that
they are accurate to $\sim 5\%$; this number depends on the region of
parameter space that is probed and larger errors can occur (M. White,
private communication).

\subsection{Constraints on $\Omega_m$ and $\sigma_8$}

Cosmological parameters are estimated by comparing the predicted
signal $m_i$ to the observed top-hat variance $d_i$ as a function of
scale $\theta_i$. We consider cold dark matter models with a flat
geometry (i.e., $\Omega_m+\Omega_\Lambda=1$). We vary the parameters
of the model, focussing on constraining the matter density $\Omega_m$
and the normalisation $\sigma_8$. For these parameters we limit the
calculations to $\Omega_m\in[0,1]$ and $\sigma_8\in[0.5,1.2]$.

\begin{figure}
\begin{center}
\epsfxsize=8.5cm
\epsffile{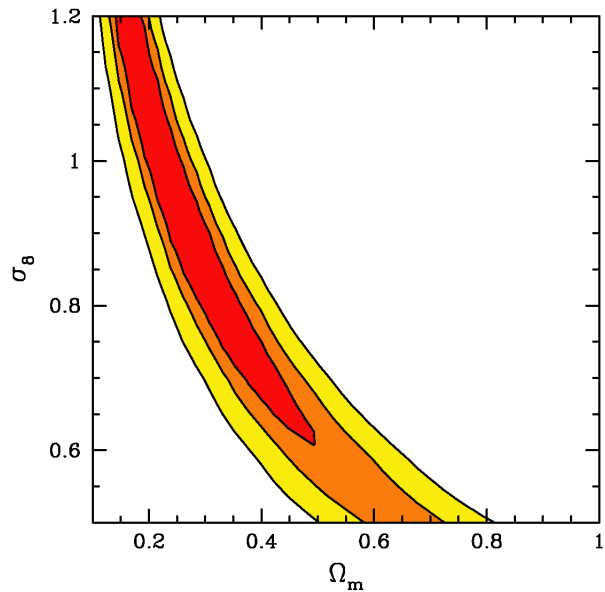}
\figcaption{\label{omsig} Joint constraints on $\Omega_m$ and
$\sigma_8$ constraints from the CFHTLS Wide data using the
Smith et al. (2003) model for the non-linear power spectrum.
The contours indicate the 68.3\%, 95.4\%, and 99.7\%
confidence limits on two paramaters jointly.
We marginalised over the Hubble parameter and source redshift
distribution as described in the text.}
\end{center}
\end{figure}

As discussed above, we vary the source redshift distribution through
$z_s=\in[0.613,0.721]$, assuming a flat prior. Furthermore, the signal
depends somewhat on the value for the Hubble parameter, for which we
use $h\in[0.6,0.8]$ as motivated by the findings of the HST Key
project (Freedman et al. 2001). In section \S5.1 we also consider $w_0$,
the dark energy equation of state. The maximum likelihood function is
given by

\begin{equation}
{\cal L}={1\over (2\pi)^n|\Cg|^{1/2}}
\exp\left[(\d_i-m_i){\Cg}^{-1}(d_i-m_i)^T\right].
\end{equation}

\noindent Here $\Cg^{-1}$ is the covariance matrix. $\Cg$ can be
decomposed as $\Cg=\Cg_n+\Cg_s$, where $\Cg_n$ is the statistical
noise and $\Cg_s$ the cosmic variance covariance matrix. The matrix
$\Cg_s$ is computed according to Schneider et al. (2002b), assuming an
effective survey area of 13.5 deg$^2$ for the CFHTLS W1 and 8.5
deg$^2$ for W3, a number density of galaxies $n_{gal}=12 {\rm \ per\
arcmin}^2$, and an intrinsic ellipticity dispersion of $\sigma_e=0.3$
per component.

\begin{figure*}
\begin{center}
\leavevmode
\hbox{%
\epsfxsize=8.5cm
\epsffile{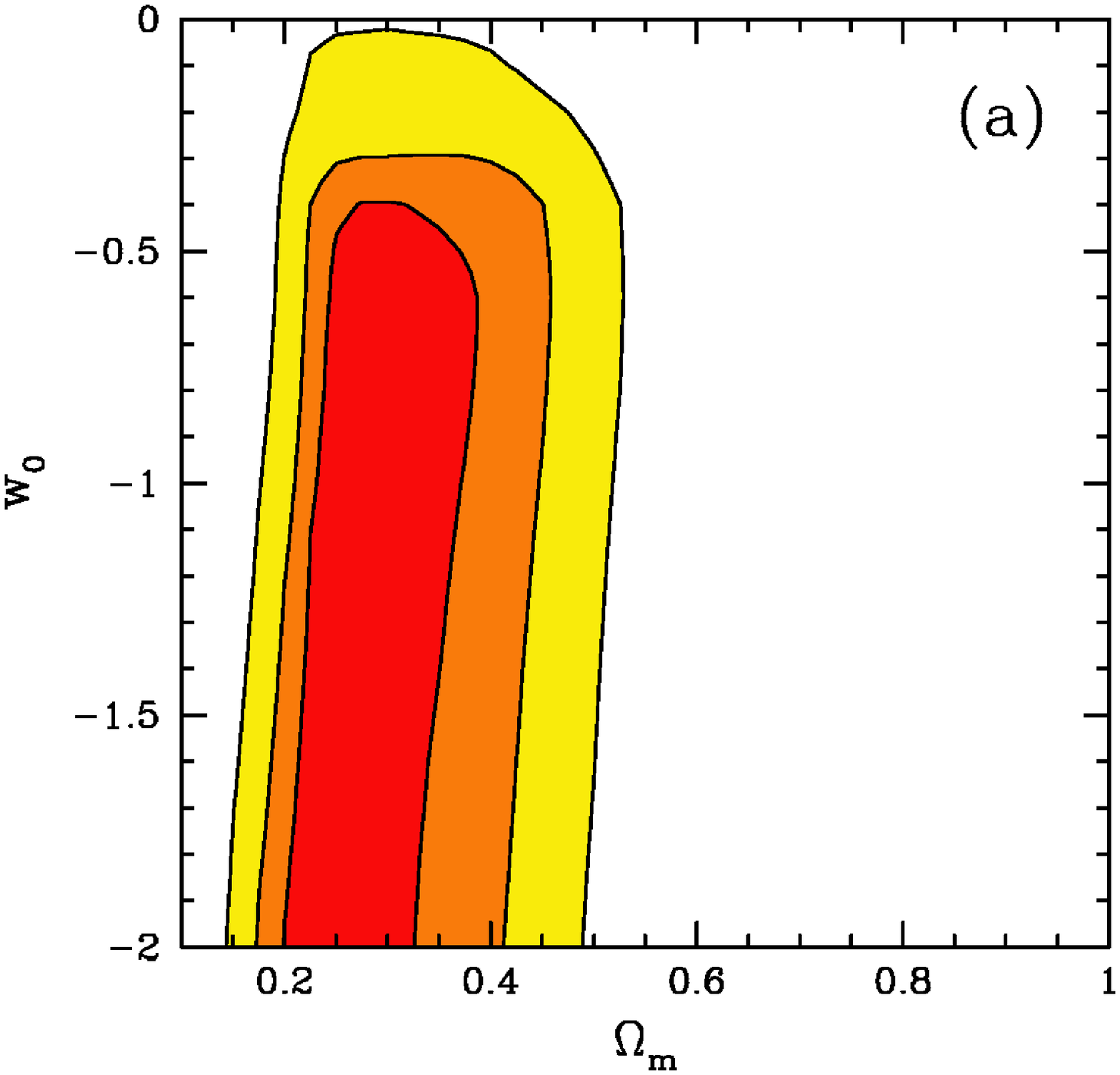}
\epsfxsize=8.5cm
\epsffile{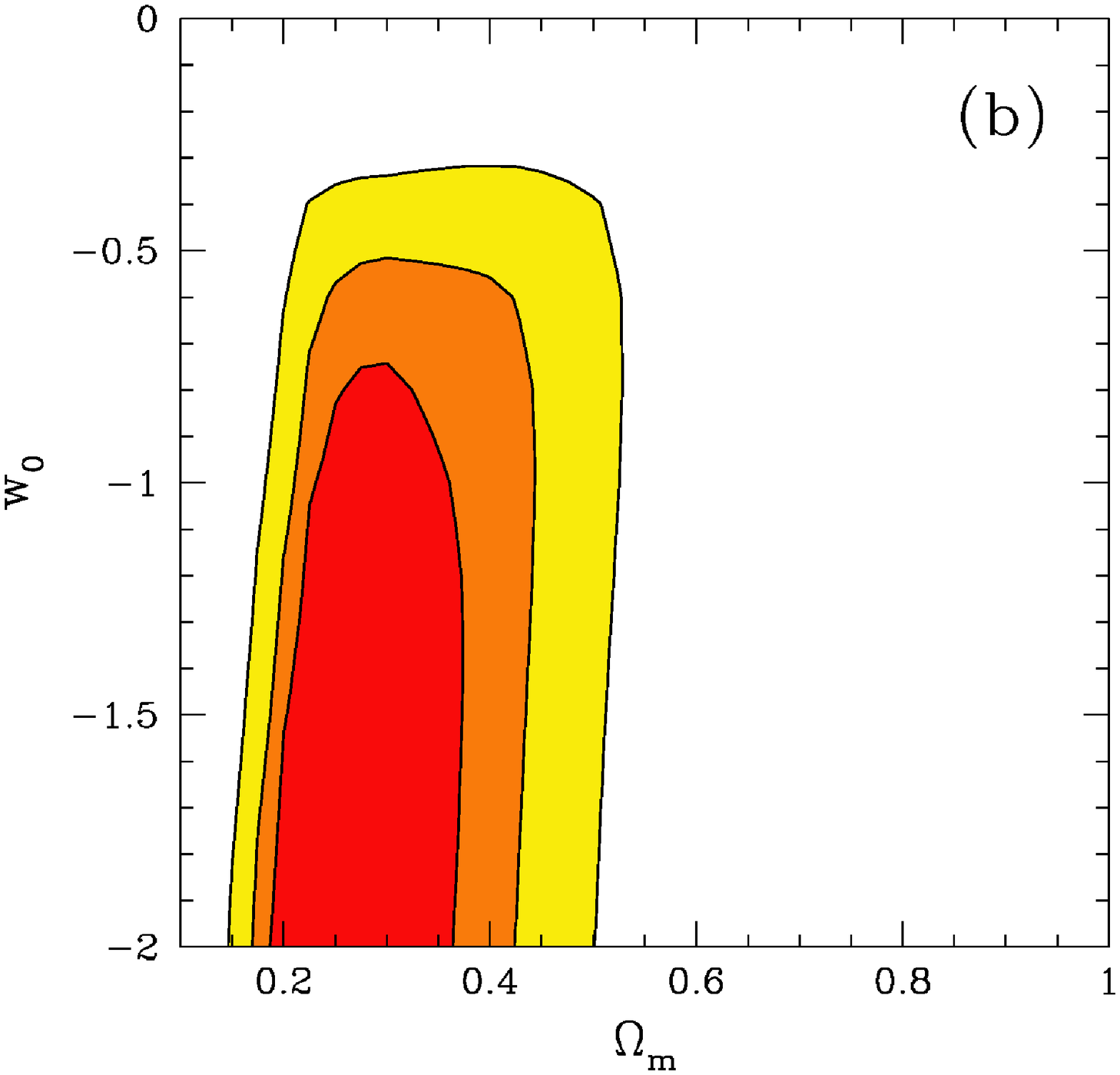}}
\figcaption{\label{eqstate} {\it panel a:} Dark energy constraints
using the measurements from the W1 and W3 fields. The contours
indicate the 68.3\%, 95.4\%, and 99.7\% confidence limits on two
paramaters jointly.  We marginalised over $\sigma_8\in[0.7,1.0]$,
$h\in[0.6,0.8]$ and the source redshift distribution as described in
the text. {\it panel b} Results when the measurements from the Deep
component (Semboloni et al. 2005) are included. We used the Peacock \&
Dodds (1996) prescription for the non-linear power spectrum.}
\end{center}
\end{figure*}

Figure \ref{omsig} shows the joint constraints on $\Omega_m$,
$\sigma_8$ using the Smith et al. (2003) model for the non-linear
power spectrum. For reference with other cosmic shear studies, we
estimate the value of $\sigma_8$ for a fiducial matter density of
$\Omega_m=0.3$. For the Peacock \& Dodds (1996) model, we obtain a
value of $\sigma_8=0.88\pm0.06$ (68\% confidence). The error includes
the statistical errors, cosmic variance and calibration errors. For
the Smith et al. (2003) model we obtain a slightly lower value of
$\sigma_8=0.85\pm0.06$. For $\Omega_m<0.4$, the degeneracy between the
two parameters is well described by $\sigma_8\propto\Omega_m^{-0.6}$.
The estimates for $\sigma_8$ based on the W1 and W3 separately also
agree well. For $\Omega_m=0.3$, we find $\sigma_8=0.87\pm0.07$ for W1
and $\sigma_8=0.75\pm0.12$ for W3. 

These estimates are in excellent agreement with published cosmic shear
results from other large surveys. Van Waerbeke et al. (2005) list a
value of $\sigma_8=0.83\pm0.07$ based on the VIRMOS-Descart survey.
Hoekstra et al. (2002c) obtained $\sigma_8=0.86^{+0.04}_{-0.05}$ from
the Red-sequence Cluster Survey. Jarvis et al. (2005) found
$\sigma_8=0.81^{+0.15}_{-0.10}$ (95\% confidence) from the CTIO
lensing survey. The constraints obtained here also agree well
with those derived from the CFHTLS Deep survey and a detailed
comparison as well as combined constraints with our measurements
are presented in Semboloni et al. (2005).

\subsection{Constraints on dark energy}

One of the major goals of the CFHTLS is the measurement of the
equation of state of dark energy. The current lack of (photometric)
redshift information seriously limits the accuracy of such a
measurement. Given the current limitations of the data, we choose a
simple model with a constant equation of state:

\begin{equation}
p=w_0 \rho.
\end{equation}

The left panel in Figure~\ref{eqstate} shows the joint constraints on
$\Omega_m$ and $w_0$ based on the measurements of the lensing signal
from the W1 and W3 fields. We marginalised over
$\sigma_8\in[0.7,1.0]$, $h\in[0.6,0.8]$ and the source redshift
distribution as described above and considered the range of
$-2<w_0<0$. To obtain these results we used the Peacock \& Dodds
fitting formula to obtain the non-linear power spectrum.  We emphasize
that there is not yet a reliable analytical fit to the non-linear dark
matter power spectrum available for non-trivial dark energy models. In
particular, the halo model proposed in Smith et al. (2004) cannot
provide an accurate description of the $w_0\ne -1$ models because the
power spectrum does not depend on $w_0$ at $z=0$. We know that a
change in $w_0$ should affect the background and therefore the change
in structure clustering, which Smith et al. (2004) do not take into
account. McDonald et al. (2005) have recently extended the halo-model
to smaller scales for $w_0\ne -1$ dark energy models. We will present
a detailed dark energy measurement in a subsequent work, taking into
account new model fitting to numerical simulations, and the CFHTLS
type Ia supernovae constraints (Van Waerbeke et al., in preparation).

Although we currently lack photometric redshift information, the
measurements of the Deep fields presented by Semboloni et al. (2005)
probe the matter power spectrum at a slightly higher mean redshift,
compared to the measurements presented here. Hence, we can improve our
constraints on $w_0$ by combining our results to those obtained from
the Deep survey (Semboloni et al. 2005). We stress that this should be
considered a consistency check rather than a complete dark energy
analysis. The result of this analysis is presented in the right panel
of Figure~\ref{eqstate}, which gives $w_0<-0.8$ with 68\% confidence.

\section{Conclusions}

We have presented the first cosmic shear analysis based on $\sim 22$
deg$^2$ (31 pointings) of deep $i'$ imaging data from the CFHT Legacy
Survey. These observations cover part of two of three survey fields
and already provide a significant increase in area compared to
previous work. These early data show a strong variation of the PSF
over the field of view.  We note that recent changes to Megaprime have
led to a significant reduction in PSF anisotropy. Nevertheless, our
early results are very encouraging as we do not detect a significant
`B'-mode, suggesting that the derived lensing signal is free of
systematics at the current level of accuracy. Comparison with an
independent pipeline (Fu et al., in preparation) shows good agreement
and demonstrates that the signal can be recovered robustly.

We assume a cold dark matter model with a flat geometry and derive
joint constraints on the matter density $\Omega_m$ and the
normalisation of the matter power spectrum $\sigma_8$, while
marginalising over the Hubble parameter and the source redshift
distribution. We consider two models to calculate the non-linear power
spectrum. For a fiducial matter density of $\Omega_m=0.3$ we find
$\sigma_8=0.88\pm0.06$ for the Peacock \& Dodds (1996)
model. Similarly, we obtain a value of $\sigma_8=0.85\pm0.06$ using
the Smith et al. (2003) approach. These estimates are in excellent
agreement with previous studies (Hoekstra et al. 2002c; Jarvis et
al. 2005; van Waerbeke et al. 2005). 

In the coming years we expect to image a total $\sim 140$ deg$^2$ in
five filters, allowing us to include photometric redshift information
for the source galaxies. This will greatly enhance our ability to
constrain cosmological parameters, most notably the equation of state
of the dark energy (e.g., see forecasts in Tereno et al., 2005).  The
measurements from Semboloni et al. (2005), based on the Deep component
of the CFHTLS probe a higher mean source redshift, thus providing
crude redshift leverage. When combining our results with the
measurements from Semboloni et al. (2005) we find $w_0<-0.8$ (68\%
confidence) based on cosmic shear measurements alone.

\acknowledgments 

We acknowledge use of the Canadian Astronomy Data Centre, which is
operated by the Dominion Astrophysical Observatory for the National
Research Council of Canada's Herzberg Institute of Astrophysics.  HH
thanks Dave Balam and Stephen Gwyn for discussions on astrometry. YM,
ES, IT and LF thank the CNRS-INSU and the ``French Programme National
de Cosmologie'' for their support to the CFHTLS cosmic shear
program. ES thanks the University of British Columbia for
hospitality. LF thanks the ``European Association for Research in
Astronomy'' training site (EARA) and the European Community for the
Marie Curie doctoral fellowship MEST-CT-2004-504604. HH, MJH and LvW
are supported by NSERC. HH and LvW acknowledge support by the the
Canadian Institute for Advanced Research (CIAR) and the Canadian
Foundation for Innovation (CFI).

\end{document}